# Tuning plasmonic field enhancement and transients by far-field coupling between nanostructures

Z. Pápa[1,2,*], J. Kasza[2], J. Budai[2], Z. Márton[2,3], Gy. Molnár[4], P. Dombi[1,2]

[1] Wigner Research Centre for Physics, Konkoly-Thege M. u. 29-33, Budapest 1121, Hungary
[2] ELI-ALPS, ELI-HU Non-Profit Ltd., Wolfgang Sandner u. 3, Szeged 6728, Hungary
[3] University of Pécs, Institute of Physics, Ifjúság u. 6, Pécs 7624, Hungary
[4] Centre for Energy Research, Institute of Technical Physics and Materials Science, Konkoly-Thege M. u. 29-33, Budapest 1121, Hungary
* corresponding author, email address: papa.zsuzsanna@wigner.hu

**Abstract**
We study how the collective effects of nanoparticles arranged in rectangular arrays influence their temporal plasmon response and field enhancement property. By systematically changing the lattice constant for arrays containing identical metal nanorods, we experimentally demonstrate how grating induced effects affect the position and, more importantly, the broadening of extinction spectra. We correlate these effects with the achievable field enhancement and the temporal duration of plasmon transients and formulate criteria for the generation of enhanced few-cycle localized plasmon oscillations.

Metallic nanostructures exhibit strong optical extinction upon interaction with light due to resonantly driven electron oscillations (plasmons). Plasmon excitation in nanostructures leads to unique phenomena such as the concentration of light to nanoscopic volumes and strong electric field enhancement [1, 2]. This field enhancement and localization are exploited in applications like surface enhanced spectroscopies, plasmonic biosensors, photovoltaics, strong-field physics or construction of nanoscale electron emitters (for a review, see [3]). Photoelectron emission from metallic nanostructures is particularly interesting since with the help of photoelectron spectra, one can experimentally determine plasmonic field enhancement factors in a direct and nondestructive manner [4-6].

For a single nanoparticle, the plasmonic resonance phenomenon depends on the particle shape, and on the dielectric properties of the particle material and the surrounding medium. In experiments, to enable a strong enough signal from these nanoparticles and to reduce discrepancies caused by defects, they are often arranged in periodic arrays where a lot of these identical nanoparticles can interact with the exciting light. For these ensembles of nanostructures, single particle plasmon resonance can be influenced by near-field or far-field electromagnetic interaction depending on the interparticle distance [7].

Nearly touching nanoparticles can directly interact via their near-fields, while nanoparticles forming an ensemble with distances exceeding those allowing near-field coupling interact via their radiating dipolar fields. First theoretical predictions on the effect of far-field coupling on the extinction properties date back to the mid-80s [8]. As the nanofabrication capabilities developed, experimental works were also published focusing on the extinction spectra of various nanoparticles arrays [9-16].

Lately, plasmonic systems preserving the temporal properties of ultrashort laser excitation along with enhanced fields have become the focus of interest. Even though the generation of few-cycle surface plasmon polariton wavepackets has been demonstrated [17, 18], it is more complicated to generate few-cycle localized plasmon oscillations with strong electric fields [19] for several reasons. Not only plasmonic nanostructures supporting large bandwidth are needed, but also femtosecond breakdown of nanoparticles has to be avoided [20]. In this paper, we will show that the behavior of individual nanorods with predesigned plasmonic properties can be tuned in a wide range by arranging them into periodic arrays with different lattice constant values. Beside the known shift of the extinction peak, here we focus on how field enhancement and temporal length of plasmon transients are affected by far-field coupling between the nanorods forming the array.

Before turning to nanoparticle arrays, we numerically investigated the extinction spectrum, the achievable field enhancement and the temporal shape of the plasmon oscillations of an individual nanorod showing resonant behavior in the NIR spectral range. These data are recorded at hot spots (corners of the nanorod, where the most prominent field concentration occurs), 0.7 nm distance from the nanorod surface. For the simulations, we applied the broad and structured spectrum of a fiber laser system delivering 9.5-fs pulses (intensity FWHM - this quantity is used for characterizing the temporal duration throughout the paper) having a spectral center-of-mass at a wavelength of 1030 nm operating at the ELI-ALPS facility (Fig. 1 (a)) [21].

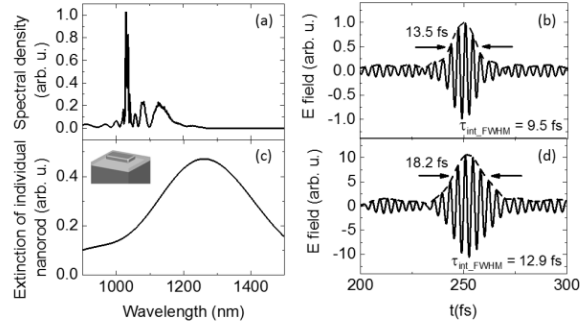

Fig. 1. Spectrum of the exciting laser pulse (a) and its temporal shape (b). Simulated single particle extinction spectrum (c), and temporal shape of plasmon transient in the hot spot upon laser excitation (d). In the bottom right corner of figs. (b) and (d) the intensity FWHM values are shown.

With the applied nanorod design (dimensions of 310 nm×120 nm×40 nm), the individual nanorod exhibits a field enhancement factor of ~10 at the corners, while the temporal length of the plasmon transient significantly exceeds that of the exciting short pulse (Fig. 1 (b)). To see how these properties are affected by far-field coupling in real samples, we measured extinction spectra of nanorod arrays with different lattice constants and compared experimental results to simulations. Details about the sample fabrication, the extinction measurement setup and the simulations are provided in supporting information.

Lattice constants of the nanorod arrays in the x and y directions ($\Lambda_x$ and $\Lambda_y$, indicated also in Fig. 2 (a)) were varied independently in a way to achieve large variations in both the position and the broadening of the extinction spectrum compared to that of the individual nanorod. For far-field coupling, extinction properties are affected by to the nanostructures' dipole fields which interfere to form collective radiation. The generated dipoles do not radiate along their axis (oriented parallel to the x direction), therefore variations in the $\Lambda_x$ parameter have a small impact on the collective behavior governed by dipolar radiation. For $\Lambda_y$, the extinction peak exhibits a redshift as the lattice constant approaches the plasmon resonance wavelength ($\lambda_{SP}$ = 1260 nm, being equal to extinction position of the individual nanostructure), since in this case, the generated local optical fields around the nanostructures are added almost in-phase between neighboring particles [9]. Furthermore, the presence of an additional, grating induced peak is expected [13], if the lattice constant exceeds a critical value. In this case, the corresponding grating order becomes radiative, and then it affects the shape of the extinction. The critical lattice constant is defined as

$$\Lambda_c = \frac{m\lambda_{SP}}{n_{subs}} \quad (1)$$

where $\lambda_{SP}$ is the wavelength of plasmons, $n_{subs}$ = 1.7 is the refractive index of the array substrate at the plasmon resonance wavelength (since the nanorods are sitting on the ITO layer, $n_{subs}$ is considered as $n_{ITO}$ at 1260 nm), and *m* is the grating order [13]. For our sample $\Lambda_c$ = 740 nm (*m*=1), which means that signatures of the grating induced band may appear above this lattice constant. To probe these collective effects, we arranged our nanorods in rectangular arrays (array size: 200 μm x 200 μm) forming two groups: in the first group $\Lambda_y$ is the same for all arrays ($\Lambda_y$=500 nm) and $\Lambda_x$ changes from 500 nm to 1100 nm by 100 nm stepsize, while in the second group $\Lambda_x$ is kept at 500 nm and $\Lambda_y$ changes between 500 nm and 1100 nm. Each nanorod forming the arrays has dimensions of 310 nm×120 nm×40 nm. Representative scanning electron microscope (SEM) images of two arrays are presented in Fig. 2 (a). Fig. 2 also shows the extinction spectra belonging to different $\Lambda_x$ and $\Lambda_y$ values, the position, the maximal value and the width of the peaks. Due to the good wavelength resolution of our extinction measurement setup (< 2 nm) and the noise-free data, we could determine these values without fitting the measured extinction spectra. The overlap of the measured and modeled spectra indicates that we can describe the properties of nanostructure arrays sufficiently in our simulations The trends exhibited are generally the same, however the calculated resonance positions are slightly red-shifted and higher in magnitude than the measured values. These discrepancies may originate from differences between the simulations and the real experimental situation. In our simulations, the rods have smooth surface and they are uniform in shape. However, the fabricated nanorods always exhibit imperfections including slightly different lengths and widths, rough sides and non-uniform heights. As a result, in the extinction measurement the average behavior of these slightly different nanorods is obtained.

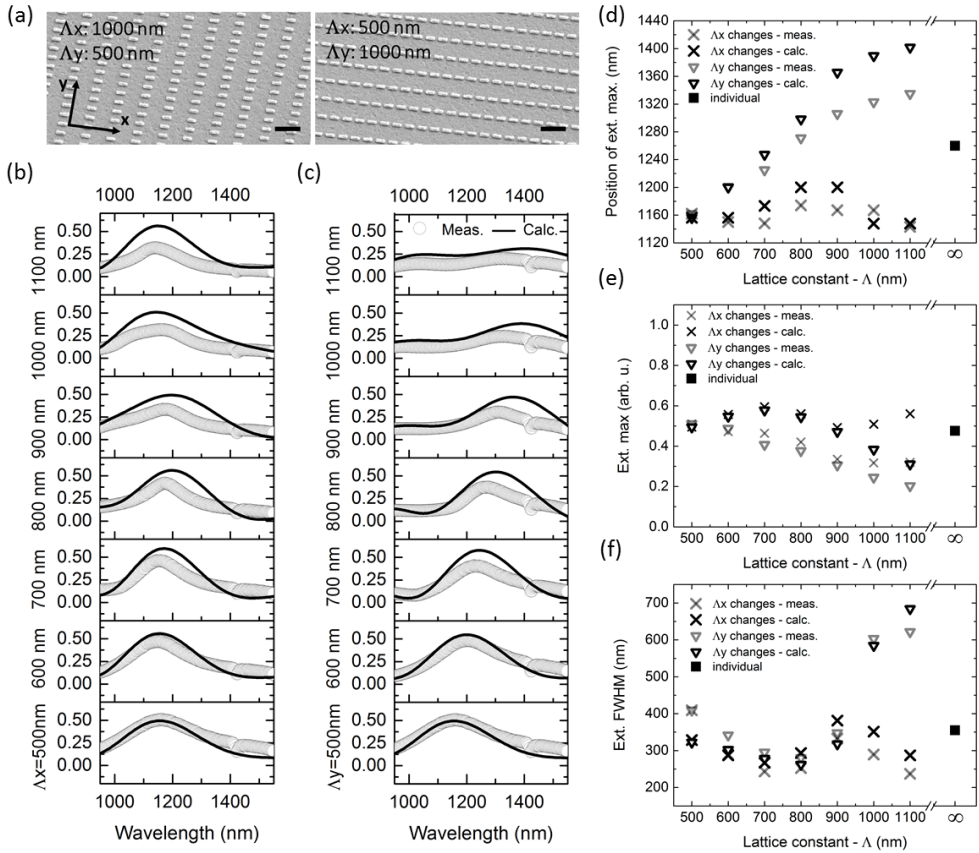

Fig. 2. (a) SEM images of representative arrays having different lattice constant values. The size of the nanorods are the same for all fields. The length of the scalebar is 1 μm. (b) and (c) Measured and calculated extinction spectra

belonging to different $\Lambda_x$ (b) and $\Lambda_y$ (c) values. Position of the extinction peaks (d), their maximal value (e) and their FWHM value (f). Crosses represent data points belonging to the $\Lambda_x$ series while open triangles denote the $\Lambda_y$ series. Square symbol at lattice constant "∞" denotes the properties of the individual nanorod.

Although the lattice constant in the x direction does not have much impact, as expected, it is still worth investigating plasmonic properties in this group of arrays. In the other group of arrays with changing $\Lambda_y$ parameter, both the redshift of the extinction peak and the broadening above the critical lattice constant parameter ($\Lambda_c$ = 740 nm) are prominent (see Fig. 2 (f)). For $\Lambda_y$ values smaller than $\Lambda_c$, the extinction spectra are blueshifted, as well as for the whole group of different $\Lambda_x$ values where $\Lambda_y$ is 500 nm. This behavior of peak positions is in agreement with the predictions of a semianalytical model [10,11] based on the coupled dipole treatment and an infinite array of particles (discussed in detail in SI). To test the effect of extinction broadening on the temporal character of plasmonic transients, we collected the time-dependent electric field (E(t)) at hot spots from the simulations (Fig. 3 (a) and (b)). Plotting these temporal durations as a function of extinction FWHM values (as shown in Fig 3 (c)) indicates that temporal shape of the plasmon transients is mainly determined by the width of the extinction spectra. This also supports that nanoparticle arrays with broadened extinction spectra can help to enable few-cycle plasmon oscillations by reducing spectral filtering of the short-pulse laser. However, nanorod arrays belonging to the symbols at the bottom left corner of Fig. 3 (c) hint at the fact that short plasmon oscillations can be obtained also with rather narrow extinction spectrum. It should be noted here, that for the extinction spectra belonging to $\Lambda_y$=800 nm and 900 nm, the appearance of the additional band around 900 nm already helps transferring spectral components without contributing to the width of the main extinction peak (not shown here). For these nanostructure arrays ($\Lambda_x$, $\Lambda_y$=800 nm and 900 nm) it is also the shift of the extinction peak, i. e. the detuning of the nanoparticle resonance that hinders the long-lived plasmon oscillations.

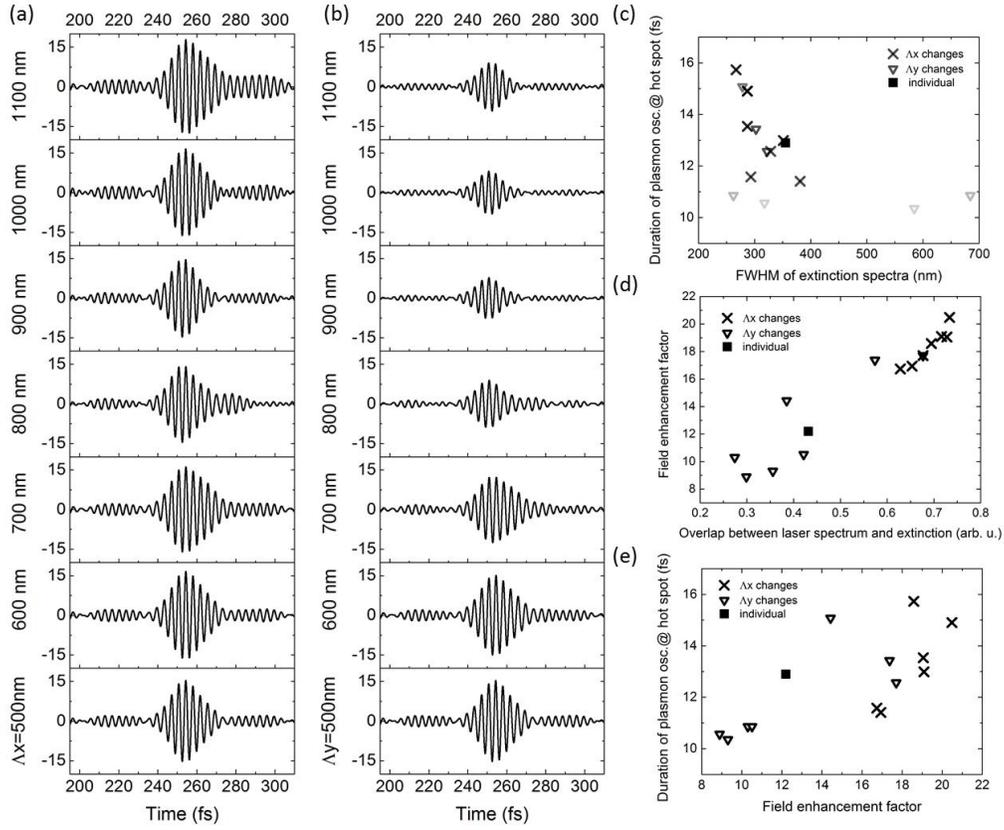

Fig. 3. (a) and (b) Calculated temporal shape of plasmon transients at the hot spot of the nanorods belonging to different $\Lambda_x$ (a) and $\Lambda_y$ (b) values. (c) Duration of plasmon transients (intensity FWHM) plotted in fig (a) and (b) as a function of the extinction FWHM value of the corresponding arrays. The darkness of the symbols hints at the field enhancement factor being equal with the maximal amplitude of the temporal signal. (d) Field enhancement factor plotted as a function of the overlap between laser spectrum and extinction curves. (e) Duration of plasmon transients at the hot spot as a function of the field enhancement factor. The square symbol represents the value belonging to an individual nanorod.

Temporal signals enable also to extract field enhancement factors being equal to the maximal amplitude of the calculated plasmon field since in the simulations the amplitude of the incoming field is unity. From the point of view of field enhancement property, the two series exhibit different behavior: for arrays with different $\Lambda_x$ values, field enhancement factors are similar, while for $\Lambda_y$, they decrease with increasing lattice constant (see the maximal field amplitudes in Fig. 3 (a) and (b)). Comparing these observations with the tendencies of Fig. 2 (e), one can draw the conclusion that for effective plasmon excitation and for large fields it is needed that extinction peaks stay close to the laser central wavelength. This criterion is also supported by the representation of Fig. 3 (d), where the field enhancement values are plotted as a function of a parameter characterizing the spectral overlap between the laser spectrum and the extinction spectra. To calculate this overlap parameter, we multiplied the laser spectrum with each extinction spectrum and integrated these product curves. The field enhancement values show a clear and monotonous increase with spectral overlap parameter emphasizing that for effective plasmon generation the spectrum of excitation and the extinction of the nanostructures should overlap as much as possible, in other words the resonance criterion has to be fulfilled.

A plausible prediction could be that the temporal duration of plasmon transients would also show a clear tendency - a decrease - with increasing spectral overlap parameter, since in this way we can reduce the spectral filtering effect exerted by the plasmonic extinction spectrum on the initially broadband laser pulse, enabling few-cycle plasmonic transients this way. However, this was not the case because really short plasmon oscillations belong to arrays with broad but redshifted extinction. Based on this, we can conclude that for short plasmon transients it is more important to transfer spectral components of the laser evenly than to match the peaks of laser spectrum and the extinction curves. Avoiding the perfect resonance also prevents the nanoparticle to sustain long-lived plasmon oscillations.

These opposite requirements, namely aiming at the overlap of the excitation spectrum and the extinction peak, and simultaneously striving for a broad and almost constant extinction spectrum make the generation of strongly enhanced but still few-cycle plasmon transients challenging. When we plot the duration of plasmon transients as a function of the field enhancement, we observe a constant increase in the duration of plasmonic field for most of the investigated arrays as the achievable near-field strength becomes larger (Fig. 3 (e)). The interesting points and the corresponding arrays are the ones located in the bottom right corner of Fig. 3 (e). For these nanostructure arrays, the field enhancement is much larger than the one obtained for individual nanorod (presented also in Fig. 3 (e)) while the temporal durations are comparable. These points belong to nanostructure arrays realizing a tradeoff between the criteria for generating strongly enhanced and short plasmon transients: they exhibit a slightly broadened and a slightly redshifted extinction compared to the individual nanoparticle. Interestingly, the most favorable configurations belong to the group of changing $\Lambda_x$, where the smaller variations in plasmonic properties are expected. Since the $\Lambda_y$ parameter has such a large impact on the position of the extinction, the overlap between the excitation spectrum and the plasmon resonance drops significantly with increasing values of $\Lambda_y$ parameter, despite the prominent broadening of the extinction due to the appearance of the grating induced band. This will deteriorate the field enhancement property of these arrays, and this way makes these configurations less favorable. For arrays with a different $\Lambda_x$ parameter, the changes are smaller both in extinction position and broadening, but there are configurations when the overlap is still large but the slight shift of the extinction peak is already enough to avoid the resonant – and therefore – long-lasting plasmon excitation. It is also important to emphasize that by having the same nanoparticle and changing only the lattice constant of the arrays, the field enhancement factor can be increased almost by a factor of two compared to the individual nanoparticle. Furthermore, by tuning the lattice constant the duration of the plasmon transient can approach the temporal length of the excitation due to the appearance of the grating induced band resulting in a broad and flat extinction. In comparison with the field enhancement property and the achievable temporal length on individual nanorods with different lengths (shown in SI), it can be concluded that tuning the lattice constant provides a more powerful tool for changing the plasmonic properties in a wide range than varying only the geometrical properties of individual nanorods.

In summary, we could demonstrate how the grating induced effects influence the position and more importantly the broadening of extinction spectra by systematically changing the lattice constant for arrays containing identical nanorods. We found that these changes of extinction have a strong impact on not only the achievable field enhancement but also on the temporal duration of plasmon transients. These observations shed light on important criteria of generating few-cycle plasmon oscillations with enhanced local fields. Based on these results, it is important to emphasize that when designing nanostructures for applications requiring both large local fields and ultrashort plasmon oscillations at the same time, it is worth considering not just the tuning possibilities offered by the properties of the individual nanostructure but also the collective effects that can be easily tuned by the lattice constants of the nanostructure arrays.

**Acknowledgements**

The authors acknowledge funding from the National Research, Development and Innovation Office of Hungary (grants VEKOP-2.3.2-16-2017-00015, 2018-1.2.1-NKP-2018-00012 and 128077), and the PETACom project no. 829153 financed by the FET Open H2020 program.

**Data Availability Statement**

The data that support the findings of this study are available from the corresponding author upon reasonable request.